\documentclass[prl,twocolumn,superscriptaddress]{revtex4}
\usepackage{graphicx,amsmath,amssymb}
\usepackage{hyperref}

\def\tr{\qopname\relax{no}{Tr}}

\def\dnabla{\stackrel{\leftrightarrow}{\nabla}}

    \def\<{\langle}         \def\>{\rangle}
      
    \def\tr{{\rm tr}}

  \def\V0{{\mathbf 0}}

\def\Ba{{\mathbf a}}  
  
 \def\Bj{{\mathbf j}} 
   
  \def\Bq{{\mathbf q}}
  \def\B0{{\mathbf 0}}

\def\be{\begin{equation}}       \def\ee{\end{equation}}
\def\bea{\begin{eqnarray}}      \def\eea{\end{eqnarray}}

\begin{document}

\title{Oscillatory vortex interaction in a gapless fermion
       superfluid  with spin population imbalance }

\author{Vladimir M. Stojanovi\'c}
\affiliation{Department of Physics, Carnegie Mellon University,
Pittsburgh, Pennsylvania 15213}
\author{W. Vincent Liu}
\affiliation{Department of Physics and Astronomy, University of
Pittsburgh, Pittsburgh, Pennsylvania 15260}
\author{Yong Baek Kim}
\affiliation{Department of Physics, University of Toronto,
Toronto, Ontario M5S 1A7, Canada} \affiliation{Department of
Physics, University of California, Berkeley, California 94720}

\date{\today}
\begin{abstract}
Using effective field theory approach we study a homogeneous
superfluid  state with a single (gapless) Fermi surface, recently
suggested as a possible phase for an ultracold Fermi gas with
spin-population imbalance. We find an unconventional form of the
interaction between vortices. The presence of gapless fermions
gives rise to an additional, predominantly attractive, potential
oscillating in space, analogous to the RKKY magnetic interaction
in metals. Our study then leads to an interesting question as to
the nature of the vortex lattice in the presence of the
competition between the usual repulsive logarithmic Coulomb and
the fermion-induced attractive oscillatory interactions.

\end{abstract}

 \maketitle

Recently a beautiful array of vortices has been observed in the
MIT experiment of ultracold Fermi gases, providing a definitive
evidence for superfluidity in this class of
systems.~\cite{Zwierlein:05} When fermionic excitations are either
fully gapped or located at a few nodal points such as in a
$d$-wave superconductor, the physics of vortices belongs to the
universality class of the XY model where the phase of the
superfluid order parameter plays the dominant role. In this case,
the vortex sector of the XY model is described by the logarithmic
Coulomb interaction between the vortices.

The development of imbalanced (spin-polarized) cold Fermi gases is
set to alter this standard picture in a fundamental fashion, with
a number of exotic superfluid states recently proposed or
revisited~\cite{Liu-Wilczek:03+05,Wu-Yip:03,Bedaque:03}. One of
those states commonly found in various theoretical
approaches~\cite{Pao+Yip:06,Sheehy:06,Son-Stephanov:06} is a
homogeneous gapless fermion superfluid with a single Fermi surface
(FS) on the molecular side of the Feshbach resonance (the BEC
regime). We will call this phase BP1 (breached-pairing state with
a single Fermi surface) after Ref.~\cite{Yi-Duan:06} (also dubbed
as ``magnetized superfluid'' { and denoted SF$_{\textrm{M}}$}
in~\cite{Sheehy:06}). Two recent experiments on the polarized
Fermi gases have brought the subject to the frontier in the cold
atom physics~\cite{expstudies06sci}. Several theoretical works
have found an analogue of the BP1 in the trap - the
superfluid-normal-mixture phase.~\cite{TrapMixedPhases:06} It is
then of great interest to examine properties of this novel
superfluid.

In { the present work}, we study the effect of gapless
fermionic excitations on the vortices in the BP1 phase by
explicitly incorporating these excitations in the low-energy
effective field theory. The resulting interaction between vortices
is no longer of the pure Coulomb form, but contains an additional
fermion-induced contribution that is attractive (between the
vortices with the same vorticity) and oscillates on a length scale
set by the spin polarization, analogous to the
Ruderman-Kittel-Kasuya-Yosida (RKKY) magnetic interaction in
metals.

The Bogoliubov-quasiparticle energy spectrum in a system of two
fermion gases with an equal mass and unequal chemical potentials
is given by~\cite{Liu-Wilczek:03+05} ($\hbar\equiv 1$)
\begin{equation}
E_{\mathbf k}^{\pm}=\sqrt{(\frac{\mathbf{k}^2}{2m}
-\mu)^2+\Delta^{2}}\pm \frac{\delta}{2} \:\:,
\end{equation}
\noindent where $\mu=(\mu_{\uparrow}+\mu_{\downarrow})/2$ is the
average chemical potential and
$\delta=\mu_{\downarrow}-\mu_{\uparrow}$ the chemical potential
mismatch. Since our treatment concerns the superfluid phase
realized on the BEC side of the Feshbach resonance, $\mu<0$ in
what follows. For definiteness, we assume that $\delta>0$. If
$\delta$ is sufficiently large (i.e., if
$\delta/2>\sqrt{\mu^2+\Delta^{2}})$, the lower branch $E_{\mathbf
k}^{-}$ of the above dispersion is gapless, with a single
effective FS. For brevity, we hereafter denote it as
$\varepsilon_{\mathbf k}$. When linearized in the vicinity of the
effective FS, this gapless dispersion adopts the generic form
$\varepsilon_{\mathbf k}=v_{b}(|\mathbf{k}|-k_{b})$, parameterized
by the ``Fermi velocity'' $v_{b}$ and { the radius
$k_{b}=[6\pi^{2}n_{b}]^{1/3}$ of the ``breached-pairing Fermi
ball'' in momentum space, where
$n_{b}=n_{\downarrow}-n_{\uparrow}$ (here
$n_{\downarrow}>n_{\uparrow}$, as follows from $\delta>0$).}

In the present context, the superfluid phase field represents the
phase of the complex Cooper pair amplitude
($\langle\psi_{\uparrow}\psi_{\downarrow}\rangle=
|\langle\psi_{\uparrow}\psi_{\downarrow}\rangle|e^{i\theta}$) and
can be separated into a (singular) vortex part and a smoothly
fluctuating field $\phi$ via $\nabla\theta(\mathbf{x},\tau)=-\Ba +
\nabla\phi(\mathbf{x}, \tau)$, where $\Ba$ is the vortex gauge
field. Let us consider a static configuration of vortex lines of
unit winding number, all parallel to the $z$-axis, with a ``vortex
charge'' density distribution
$\rho(\mathbf{x})=2\pi\sum_{\alpha}\delta^{(2)}(\mathbf{x}-
\mathbf{x}_{\alpha})$. The gauge field $\Ba$ is related to $\rho$
by the standard equation $\nabla
\times\mathbf{a}=-\kappa_{0}\rho(\mathbf{x})\hat{\mathbf{e}}_{z}$,
where $\kappa_{0}=\pi\hbar/m$ ($\hbar$ restored for clarity) is
the circulation quantum. In momentum space this reads
$\mathbf{q}\times \mathbf{a}_{\mathbf{q}}=
i\kappa_{0}\tilde{\rho}(\mathbf{q})\hat{\mathbf{e}}_{z}$, where
$\mathbf{q}$ is a two-dimensional wave-vector
($\mathbf{q}\cdot\hat{\mathbf{e}}_{z}=0$) of the phase
fluctuations and $\tilde{\rho}(\mathbf{q})$ the Fourier transform
of $\rho(\mathbf{x})$. For convenience, we adopt the Coulomb gauge
$\nabla\cdot \mathbf{a} =0$ in which the vector field
$\mathbf{a}_{\mathbf{q}}$ is purely transverse ($\Bq\cdot
\mathbf{a}_{\mathbf{q}}=0$).

We start from a low-energy effective  Lagrangian for the gapless
branch of the Bogoliubov quasiparticles (described by the field
$\chi(x)$) and the superfluid phase field. This Lagrangian obeys
two global $U(1)$ symmetries, one of which corresponds to the
total atom number conservation ($U_{c}(1)$), and the other one to
the conservation of spin population imbalance ($U_{s}(1)$).  Son
and Stephanov~\cite{Son-Stephanov:06} first constructed an
effective Lagrangian of this kind for the case of small gapless
fermion density (corresponding to small spin population imbalance)
to obtain a phase diagram. Here, we extend their approach to  an
arbitrary spin population imbalance.~\cite{son:06remark}In the
imaginary-time path-integral formalism, our effective Lagrangian
reads
\begin{equation}\label{lagrang}
\begin{array}{rl}
\mathcal{L}=&\displaystyle \chi^{*}[\partial_{\tau}
+\varepsilon(-i\nabla)]\chi+c_{1}(\partial_{\tau}\theta)^{2}
+c_{2}(\nabla\theta)^{2} \\
&\displaystyle+c_{3}\chi^{*}\chi\Big[i\partial_{\tau}\theta +
\frac{1}{2m_{p}}(\nabla\theta)^{2}\Big]+\nabla\theta\cdot\mathbf{j}
+\cdots \:\:,
\end{array}
\end{equation}
where `$\cdots$' stands for higher-order derivative terms of the
$\theta$ field;
$\mathbf{j}=(\chi^{*}\nabla\chi-\nabla\chi^{*}\chi)/(2m_{p}i)$,
where $m_{p}=2m$ is the mass of the Cooper pair;
$\varepsilon(-i\nabla)$ is the gapless fermion dispersion, written
in the coordinate representation. The phenomenological parameters
$c_{1}$, $c_{2}$, and $c_{3}$ are not constrained by the $U(1)$
symmetries, but can be determined from the Galilean-invariance of
the fermion-independent part of the above Lagrangian ;
$c_{1}=\partial n/\partial \mu$ (where $n = n_{\uparrow} +
n_{\downarrow}$), while $c_{2}$ and $c_{3}$ are related to the
superfluid density in a manner to be described shortly. Under
Galilean boost with velocity $\mathbf{u}$ we have  the following
transformation properties : $\partial_{\tau}\rightarrow
\partial'_{\tau'}=\partial_{\tau}- i(\mathbf{u}\cdot\nabla)
\:,\: \nabla\rightarrow\nabla'=\nabla\:,\:\chi(\mathbf{x},\tau)
\rightarrow\chi'(\mathbf{x'},\tau')=\chi(\mathbf{x},\tau)\:,
\:\theta(\mathbf{x},\tau)\rightarrow
\theta'(\mathbf{x'},\tau')=\theta(\mathbf{x},\tau)-m_{p}\mathbf{u}
\cdot\mathbf{x}-\frac{i}{2}m_{p}\mathbf{u}^{2}\tau$. An
alternative effective theory, formulated in terms of
majority-species original fermions rather than Bogoliubov
quasiparticles, is given in Ref.~\cite{nishida+son:06}.

Because the Lagrangian has to be invariant under the $U_{c}(1)$
particle number symmetry $\theta\rightarrow \theta+\alpha$, it
contains the coordinate and time derivatives of $\theta$, but not
$\theta$ itself. Recast in terms of $\phi$ and $\mathbf{a}$, it
adopts the form
\begin{equation}\label{lagr}
\begin{array}{rl}
\mathcal{L}=&\displaystyle
\chi^{*}[\partial_{\tau}+\varepsilon(-i\nabla)]
\chi+c_{1}(\partial_{\tau}\phi)^{2}+c_{2}(\nabla\phi
-\mathbf{a})^{2}\\
&\displaystyle+c_{3}\chi^{*}\chi\Big[i\partial_{\tau}\phi+
\frac{(\nabla\phi-\mathbf{a})^{2}}{2m_{p}}\Big]
+(\nabla\phi-\mathbf{a})\cdot\mathbf{j} \:\:.
\end{array}
\end{equation}

We derive an effective phase-only action $S[\theta]\equiv
S[\phi,\mathbf{a}]$ by integrating out the fermionic degrees of
freedom: $e^{-S[\theta]}=\int
D(\chi^{*},\chi)e^{-S[\chi,\theta]}$, where
$S[\chi,\theta]=\int_{0}^{\beta}d\tau \int d\mathbf{x} \
\mathcal{L}$ is the Euclidean action corresponding to the last
Lagrangian ($\beta\equiv (k_{B}T)^{-1}$). In order to accomplish
this, we first note that the fermion field enters the Lagrangian
\eqref{lagr} through a quadratic form
$\chi^{*}K\chi=\chi^{*}(-\mathcal{G}_{0}^{-1}+X)\chi$, where
$\mathcal{G}_{0}=[-\partial_{\tau}-\varepsilon(-i\nabla)]^{-1}$ is
the noninteracting fermion propagator, and $X=X^{(1)}+X^{(2)}$
where $X^{(1)}=c_3
i\partial_{\tau}\phi+\frac{1}{2m_{p}i}(\nabla\phi-\mathbf{a})\cdot
\dnabla$ and
$X^{(2)}=\frac{c_{3}}{2m_{p}}(\nabla\phi-\mathbf{a})^{2}$ are
respectively of the first and second order in fields $\phi$ and
$\mathbf{a}$. Integrating out the $\chi$ fields gives rise to the
fermion contribution to the action $S [\phi,\mathbf{a}]$:
\begin{equation}\label{expan}
S_{F}[\phi, \mathbf{a}]=-\tr\ln K =
\textrm{const.}+\sum_{n=1}^{\infty}\frac{1}{n}
\tr[(\mathcal{G}_{0}X)^{n}]\:\:.
\end{equation}
\noindent { We expand \eqref{expan} to first order in $X^{(2)}$
(tree level) and to second order in $X^{(1)}$ (one-loop order).
The effective phase-only action $S[\phi,\mathbf{a}]$ is then
obtained by gathering $S_{F}[\phi,\mathbf{a}]$ and the
fermion-independent terms of the original action. In
momentum-frequency space ($q \equiv (\mathbf{q},i\omega_{l})$)
\begin{equation}
\begin{array}{rl}
&\displaystyle S[\phi,\mathbf{a}]=\sum_{q}\Bigg\{
\left(c_{2}+\frac{n_{b}}{2m_{p}}\:c_{3}\right)
\mathbf{q}^{2}+\frac{1}{2m_{p}^{2}}R_{ij}(q)q_{i}q_{j} \\
&\displaystyle +\left(c_{1}-\frac{\Pi(q)}{2}\right)
\omega_{l}^{2}\Bigg\}\phi_{q}\phi_{-q} \\
&\displaystyle +\sum_{q}\Bigg\{c_{2} +\frac{n_{b}}{2m_{p}}\: c_{3}
+\frac{P(q)}{2m_{p}^{2}}\Bigg\}{\mathbf{a}}_{q}
\cdot{\mathbf{a}}_{-q} \:\:\:. \label{phasef1}
\end{array}
\end{equation}
\noindent where the first and second terms correspond to the
propagating Goldstone mode and the topological vortex part of the
broken $U_{c}(1)$ symmetry, respectively. [Summation over repeated
indices in the last equation is implicit.] Here $\Pi(q)$ is the
fermion density polarization bubble, { while
\begin{eqnarray}
R_{ij}(q) &=& \frac{1}{\beta V}
\sum_{k}\mathcal{G}_{0}(k)\mathcal{G}_{0}(k+q)
(k_{i}+\frac{q_{i}}{2})(k_{j}
+\frac{q_{j}}{2})  \\
P(q) &=& \frac{1}{\beta V}\sum_{k}\mathcal{G}_{0}(k)
\mathcal{G}_{0}(k+q)\:\frac{\mathbf{k}_{\perp}^{2}}{2}
\label{pqdef}
\end{eqnarray}
\noindent represent the longitudinal and transverse
current-current correlation functions, respectively, with
$\mathbf{k}_{\perp}$ being the transverse component of the
three-dimensional vector $\mathbf{k}$ with respect to
$\mathbf{q}$.}}

It is important to point out that there is no RPA-type correction
from the interaction vertex $\Bj \cdot \nabla\phi$ to the
transverse current-current correlation function; this is manifest
in our choice of the Coulomb gauge for the topological gauge field
${\bf a}$.

The magnitude of wave-vector $\mathbf{q}$ of the phase
fluctuations has an upper cutoff of order
$k_{\Delta}=(2m\Delta)^{1/2}$, a momentum scale corresponding to
the pairing gap. This choice of momentum cutoff is a consequence
of the BP$1$ phase being realized in the strong-coupling regime on
the BEC side of the Feshbach resonance, where $\Delta$ is related
to the binding energy of a Feshbach molecule. It is also important
to emphasize that typical $|\mathbf{q}|$ is not necessarily small
with respect to $k_{b}$; the latter is controlled by the spin
imbalance and can be arbitrarily small.

It is straightforward to show that $R_{ij}(q)= R(q)\delta_{ij}$.
Consequently, the phase-only action in Eq.~\eqref{phasef1} in the
zero-temperature static limit reduces to
\begin{equation}\label{phasef2}
\begin{array}{rl}
S[\phi,\mathbf{a}]=& \displaystyle\sum_{\mathbf{q}}
\left(c_{2}+\frac{n_{b}}{2m_{p}}\:c_{3}
+\frac{R^{0}_{\mathbf{q}}}{2m_{p}^{2}}\right)\mathbf{q}^{2}
\phi_{\mathbf{q}}\phi_{-\mathbf{q}} \\
&\displaystyle+\sum_{\mathbf{q}}\left(c_{2}+\frac{n_{b}}{2m_{p}}\:c_{3}
+\frac{P^{0}_{\mathbf{q}}}{2m_{p}^{2}}\right)
{\mathbf{a}}_{\mathbf{q}}\cdot{\mathbf{a}}_{-\mathbf{q}} \:\:,
\end{array}
\end{equation}
\noindent where $P^{0}_{\mathbf{q}}$ and $R^{0}_{\mathbf{q}}$ are
the zero-temperature static limits of $P(q)$ and $R(q)$,
respectively.

{ The superfluid mass density $\rho_{s}=n_{s} m_{p}$, which
plays the role of rigidity in the present problem (``spin-wave''
stiffness in the XY-model terminology), can be identified from the
long-wavelength ($\mathbf{q}\rightarrow 0$) limit through the
relation
\begin{equation}\label{}
\frac{\rho_{s}}{2}=c_{2}+\frac{n_{b}}{2m_{p}}\:c_{3}+
\frac{R^{0}_{\mathbf{q}=0}}{2m_{p}^{2}} \:\:.
\end{equation}
\noindent This equation constraints the coefficients $c_{2}$ and
$c_{3}$, which reduces the phase-only action in Eq. \eqref{phasef2}
to
\begin{equation}\label{phasef3}
\begin{array}{rl}
S[\phi,\mathbf{a}]=&\displaystyle \sum_{\mathbf{q}}
\frac{n_{s}}{2m_{p}}\:\mathbf{q}^{2}\phi_{\mathbf{q}}
\phi_{-\mathbf{q}}  \\
&\displaystyle +\sum_{\mathbf{q}}\left(\frac{n_{s}}{2m_{p}}
+\frac{P^{0}_{\mathbf{q}}-R^{0}_{\mathbf{q}=0}}{2m_{p}^{2}}
\right){\mathbf{a}}_{\mathbf{q}}\cdot{\mathbf{a}}_{-\mathbf{q}}
\:\:,
\end{array}
\end{equation}
\noindent which is free of the phenomenological parameters of the
original theory.}

Integrating out the phase field $\phi$ in the action
\eqref{phasef3}, taking account of the fact that
$R^{0}_{\mathbf{q}=0}=P^0_{\mathbf{q}=0}$,  leads to the following effective action for
vortices
\begin{equation}\label{rezultat}
S_\mathrm{eff}[\mathbf{a}]=\sum_{\mathbf{q}}
\left(\frac{n_{s}}{2m_{p}}+\frac{P^{0}_{\mathbf{q}}
-P^{0}_{\mathbf{q}=0}}{2m_{p}^{2}}\right)\mathbf{a}_{\mathbf{q}}
\cdot\mathbf{a}_{\mathbf{-q}}  \:\:.
\end{equation}
\noindent (Since the vortex gauge field belongs to the classical
sector of the theory, the derived effective action contains only
the $\omega_{l}=0$ part). The last result, combined with the
identity
$\tilde{\rho}(\mathbf{q})\tilde{\rho}(\mathbf{-q})=(\mathbf{q}^{2}
/\kappa_{0}^{2})(\mathbf{a}_{\mathbf{q}}\cdot
\mathbf{a}_{\mathbf{-q}})$, yields the effective vortex
interaction potential:
\begin{equation} \label{effpot}
V_\mathrm{eff}(\mathbf{q})=\kappa_{0}^{2}
\left(\frac{n_{s}}{2m_{p}}\frac{1}{\mathbf{q}^{2}}
+\frac{1}{2m_{p}^{2}}\frac{P^{0}_{\mathbf{q}}
-P^{0}_{\mathbf{q}=0}}{\mathbf{q}^{2}} \right)\:\:.
\end{equation}
\noindent Thus besides the conventional long-range component
proportional to $1/\mathbf{q}^{2}$, we have an additional
component
\begin{equation}\label{vind}
V_\mathrm{ind}(\mathbf{q})=\frac{\kappa_{0}^{2}}{2m_{p}^{2}}
\frac{P^{0}_{\mathbf{q}}-P^{0}_{\mathbf{q}=0}}{\mathbf{q}^{2}}
\end{equation}
\noindent induced by the gapless fermions.

\begin{figure}[htbp]
\begin{center}
\includegraphics[width=0.8\linewidth]{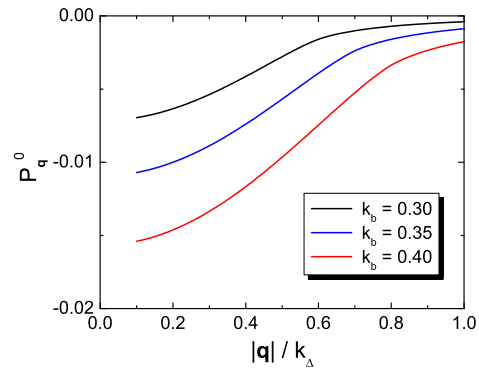}
\end{center}
\caption{Transverse current response function $P_{\mathbf{q}}^{0}$
as a function of dimensionless momentum, for $m=1.0$ and
$\Delta/|\mu|=2.0$. Values of $k_{b}$ are given in units of
$k_{\Delta}$.} \label{fig:transvcurrent}
\end{figure}

Response function $P^{0}_{\mathbf{q}}$ has to be calculated
numerically. Yet, prior to numerical computations,
$P^{0}_{\mathbf{q}}$ can be reduced to a two-dimensional
principal-value integral
\begin{equation}\label{twodint}
P^{0}_{\mathbf{q}}=\frac{k_{b}^{3}}{(2\pi)^{2}}\mathcal{P}
\int_{0}^{1}|\mathbf{k}|^{4}d|\mathbf{k}|
\int_{0}^{\pi}\frac{1-\cos^{2}\theta}{\varepsilon_{\mathbf{k}}-
\varepsilon_{\mathbf{k+q}}}\sin\theta d\theta  \:\:,
\end{equation}
\noindent where momenta $\mathbf{k}$ and $\mathbf{q}$ are
expressed in units of $k_{b}$. For small $\mathbf{q}$
($|\mathbf{q}|<0.1 k_{\Delta}$), numerical evaluation becomes
troublesome because of the strongly singular character of the
integrand in Eq.~\eqref{twodint}. However, in the limiting case
$|\mathbf{q}|\rightarrow 0$, by replacing dispersion
$\varepsilon_{\mathbf{k}}$ with its linearized form
($\varepsilon_{\mathbf{k}}\rightarrow v_{b}(|\mathbf{k}|-k_{b})$)
we can derive the analytical result $P^{0}_{\mathbf{q}}
\rightarrow -k_{b}^{4}/(6\pi^{2}v_{b})$. Since
$v_{b}=\left|\partial\varepsilon_{\mathbf k}/
\partial\mathbf{k}\right|_{|{\mathbf k}|=k_{b}}$,
the last result can be expressed as
\begin{equation}\label{}
P^{0}_{\mathbf{q}} \rightarrow -\frac{mk_{b}^{3}}{6\pi^{2}}
\frac{\sqrt{(\frac{k_{b}^2}{2m}-\mu)^2+\Delta^{2}}}
{\frac{k_{b}^{2}}{2m}-\mu} \qquad (\: |\mathbf{q}|\rightarrow 0
\:) \:\:.
\end{equation}
\noindent Some typical results of numerical evaluation of the
response function $P^{0}_{\mathbf{q}}$ for $|\mathbf{q}|\geq 0.1
k_{\Delta}$ are displayed in Fig. \ref{fig:transvcurrent} (where
$2k_{b}<k_{\Delta}$). The salient characteristic of these results
is a knee-like feature at $|\mathbf{q}|=2k_{b}$, which reflects
the existence of an effective FS of diameter $2k_{b}$. It bears
analogy to the $2k_{F}$-feature of the paramagnetic spin
susceptibility in $3$D, responsible for the RKKY indirect-exchange
interaction in metals, albeit the $2k_b$-feature found here comes
from the current-current correlator so that it is not directly
related to the RKKY interaction. The values of
$P^{0}_{\mathbf{q}}$ obtained analytically in
$|\mathbf{q}|\rightarrow 0$ limit differ just slightly from
numerical values at $|\mathbf{q}|= 0.1 k_{\Delta}$, indicating
that $P^{0}_{\mathbf{q}}$ can be approximated as a constant in
this numerically-inaccessible region $0<|\mathbf{q}|< 0.1
k_{\Delta}$.

By exploiting the rotational invariance of the induced potential
in momentum space, in real space we obtain
\begin{equation}\label{}
V_\mathrm{ind}(r)=\frac{1}{2\pi}\int_{0}^{k_{\Delta}}
|\mathbf{q}|V_\mathrm{ind}
(|\mathbf{q}|)J_{0}(|\mathbf{q}|r)d|\mathbf{q}| \:\:,
\end{equation}
\noindent where $J_{0}(x)$ is the zeroth-order Bessel function of
the first kind. { Our numerical calculations for different
values of relevant parameters ($k_{b},\Delta$) show that the
induced potential has RKKY-like oscillating behavior, with
attractive character ($dV_\mathrm{ind}/dr>0$) at short distances.
To calculate the effective vortex interaction potential, given by
the sum of $V_\mathrm{ind}$ and the repulsive logarithmic part
$V_{0}(r)=-(\kappa_{0}^{2}n_{s}/2m_{p})\ln(k_{\Delta}r)$, we have
computed the superfluid density by adapting the method of
Ref.~\cite{he:06} to the special case of a single gapless FS.

In order to elucidate the realm of validity of our effective
theory and make contact with experiments, it is useful to estimate
the physical healing length $\xi=(8\pi n_{s} a_{m})^{-1/2}$
($a_{m}$ being the molecular scattering length), with the inverse
of the momentum scale $k_{\Delta}$. It is known that in the
strong-coupling BEC regime of a superfluid Fermi gas with equal
populations of two hyperfine spin components the molecular
scattering length is given by $a_{m}=0.6\:a_{f}$ ($a_{f}$ being
the scattering length between fermionic atoms).~\cite{Petrov+:04}
For a polarized Fermi gas, as shown by Sheehy and Radzihovsky,
$a_{m}$ decreases monotonously as a function of the polarization
and vanishes at the boundary of first-order phase transition to
the phase separated state. Therefore, right at the boundary to
phase separation and in the immediate vicinity of it the coherence
length becomes much greater than $k_{\Delta}^{-1}$, thus making
the quantitative implications of our theory not directly
applicable in this special case. Taking $a_{m}(P=0)$ in place of
$a_{m}$, together with typical values of $n_{s}/n$ and
$\Delta/\epsilon_{F}$ ($\epsilon_{F}=k_{F}^{2}/(2m)$, where
$k_{F}=(3\pi^{2}n)^{1/3}$ is the momentum scale set by the total
fermion density) in the BEC regime, we estimate that $\xi$ is of
the same order as $k_{\Delta}^{-1}$ when $|\kappa|\sim 1$-$100$.
By taking into account the fact that $\xi/k_{\Delta}^{-1}\propto
(a_{m}/a_{f})^{-1/2}$, we can infer that the above estimate is
just slightly modified as a result of $a_{m}$ decreasing in the
vicinity of the phase separation line. }

\begin{figure}[thb]
\begin{center}
\includegraphics[width=0.7\linewidth]{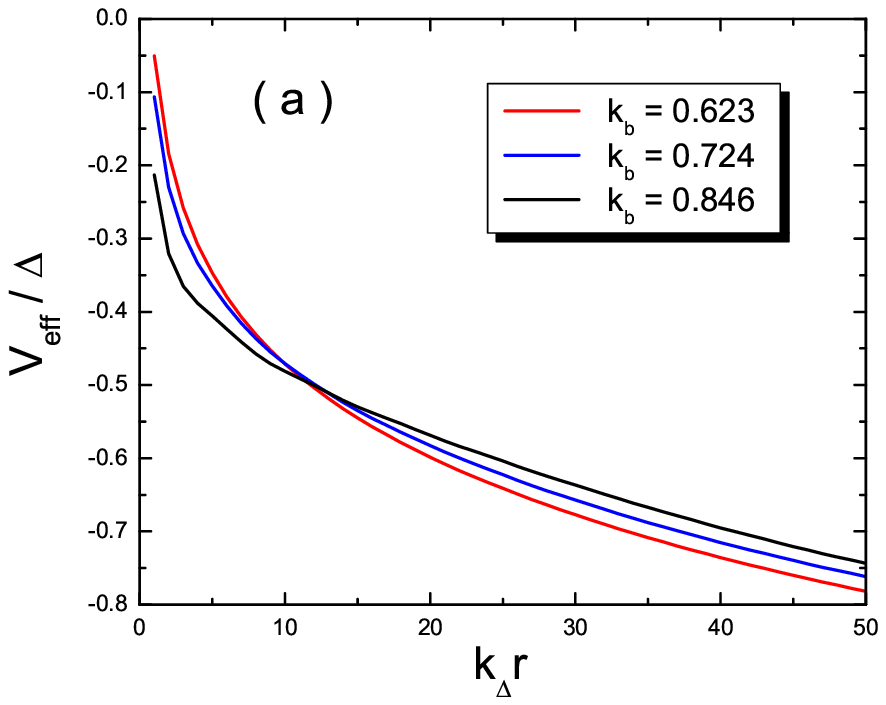}
\includegraphics[width=0.7\linewidth]{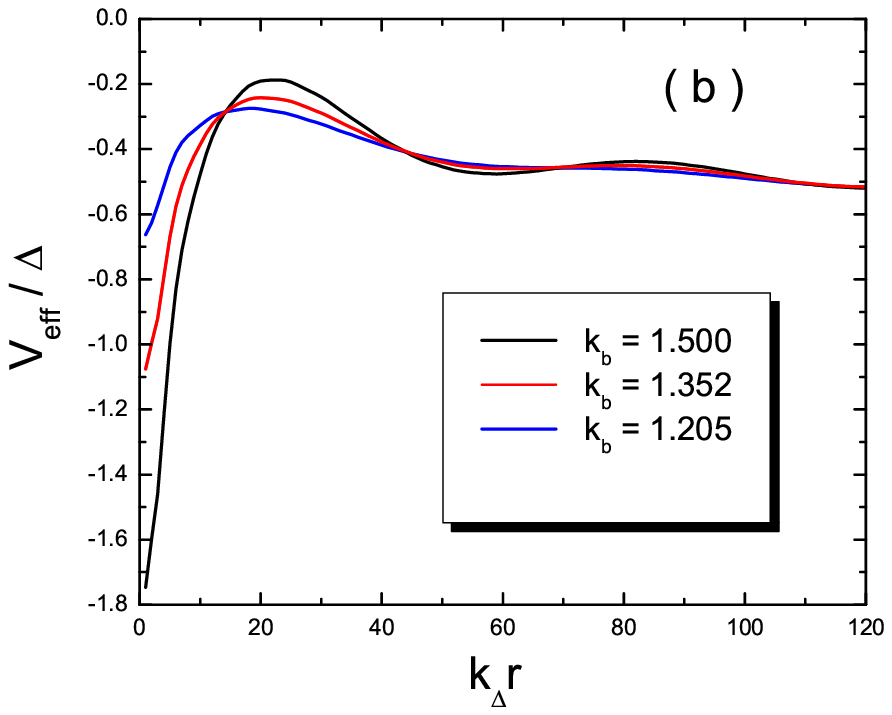}
\end{center}
\caption{Effective vortex interaction potential in real space (in
units of $\Delta$), for $m=1.0$ and $\Delta/|\mu|=2.0$. Values of
$k_b$ in both plots are expressed in units of $k_\Delta$: (a)
$k_{b}$ = 0.623; 0.724; 0.846 correspond to polarizations $P$ =
0.155; 0.227; 0.314, and (b) $k_{b}$ = 1.500; 1.352; 1.205
correspond to polarizations $P$ = 0.763; 0.702; 0.626,
respectively.}\label{fig:vortex_int}
\end{figure}

\begin{figure}[htb]
\begin{center}
\includegraphics[width=0.7\linewidth]{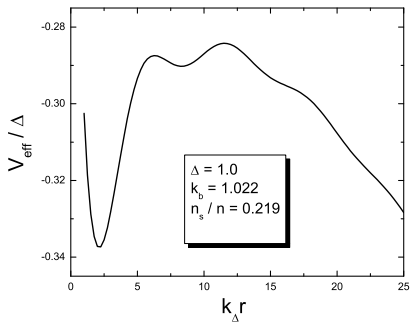}
\end{center}
\caption{Effective vortex interaction potential in real space (in
units of $\Delta$), for $m=1.0$ and $\Delta/|\mu|=1.0$.
$k_{b}=1.022$ (in units of $k_{\Delta}$) corresponds to
$P=0.504$.}\label{fig:effpot3}
\end{figure}

{ As our calculations demonstrate, the effective vortex-vortex
interaction shows three characteristic types of behavior, i.e.
three polarization-dependent regimes. The critical polarizations
corresponding to the boundaries between these different regimes
are not universal but depend on the actual location in the part of
the phase diagram pertaining to the BP1 phase.

In the regime of relatively low polarization, the total potential
is dominated by the conventional repulsive logarithmic part; the
effective vortex interaction is repulsive
($dV_{\textrm{eff}}/dr<0$) at all distances. The resulting vortex
phase is accordingly expected to be conventional, with triangular
vortex arrangement. An example is shown in
Fig.~\ref{fig:vortex_int}a.

In the other extreme - the regime of high polarization, the
induced potential plays a dominant role at short and intermediate
distances. This renders the total potential attractive at short
distances, with pronounced oscillating features resembling the
RKKY interaction, as illustrated in Fig.~\ref{fig:vortex_int}b.

The attractive nature of two-body interaction already at short
distances suggests an instability of the vortex lattice. However,
whether this instability really occurs is still an open question
for the following reasons. The physics at distances shorter than
the healing length $\xi$ (to be discussed in the next section) is
not captured by our effective theory ; also, the multi-vortex
interactions, not considered here but certainly allowed as higher
orders in the effective vortex action, may support unusual vortex
phases.

Apart from these two extreme regimes, in a narrow window of
parameters the total potential is repulsive at short distances
($r\approx (2-3)k_{\Delta}^{-1}$) and becomes attractive at
intermediate ones (Fig.~\ref{fig:effpot3}).

Due to the finite range of the RKKY-like induced potential, the
truly long-distance dependence of the effective potential (for all
polarizations) is governed by the infinite-range repulsive
logarithmic interaction.}


In summary, we have found that the vortex interaction potential
has a remarkable form in the gapless fermion superfluid (BP1), a
recently proposed phase for a resonantly interacting imbalanced
Fermi gas in the BEC regime. The vortex interaction found here
contains an attractive fermion-induced potential with the
RKKY-like oscillating character, in addition to the usual
repulsive Coulomb potential. In the conventional superfluids, the
point vortices with the repulsive logarithmic ($2$D Coulomb)
potential would form the triangular lattice.~\cite{campbell:89}
Our finding of the novel vortex interaction, therefore, opens up
an intriguing question about the vortex lattice structure in the
spin-imbalanced gapless fermion superfluid state.

{\it Acknowledgments} : We thank D. T. Son for useful discussions.
V.M.S. and W.V.L. were supported in part by ORAU and ARO
(W911NF-07-1-0293). Y.B.K. was supported by the NSERC, CRC, CIAR,
KRF-2005-070-C00044, and the Miller Institute for Basic Research
in Science at UC Berkeley through the Visiting Miller
Professorship. This research was supported during the completion
in part in KITP at UCSB by the National Science Foundation under
Grant No. NSF PHY05-51164.

\bibliography{imbalance_fermi}

\begin{thebibliography}{15}
\bibitem[{son()}]{son:06remark}
\bibinfo{note}{They discussed only the case of small spin imbalance and
  accordingly discarded the symmetry-allowed term with coefficient $c_{3}$ in
  Eq.~(2). Note also that their convention for the superfluid phase field is
  different from ours in two aspects : they defined it as half of the phase of
  the complex Cooper pair amplitude and with an extra minus sign.}
\end{thebibliography}
\bibliographystyle{apsrev}

\end{document}